\documentclass[10pt]{iopart}

\usepackage{epsfig}
\usepackage{graphics}
\usepackage{graphicx}
\usepackage{bm}
\usepackage{color}
\usepackage{hyperref}
\usepackage{amssymb}
\usepackage{longtable}
\usepackage{rotating}
\usepackage{color}
\usepackage{epsf}
\usepackage{fancyhdr}

\definecolor{purple}{rgb}{0.5,0,0.5}

\def\bea{\begin{eqnarray}}
\def\eea{\end{eqnarray}}

\newcommand{\be}{\begin{equation}}
\newcommand{\ee}{\end{equation}}
\newcommand{\ber}{\begin{eqnarray}}
\newcommand{\eer}{\end{eqnarray}}
\newcommand{\Mc}{{\cal M}}
\newcommand{\cm}{{\cal M}}

\newcommand{\Ms}{M_{\odot}}
\newcommand{\ie}{i.e. }

\def\Gpc{\mathrm{Gpc}}

\def\cm{\mathcal{M}}

\def\Ms{M_{\odot}}
\def\inst{\textrm{\mbox{\tiny{inst}}}}
\def\conf{\textrm{\mbox{\tiny{conf}}}}

\def\FWF{\textrm{\mbox{\tiny{FWF}}}}
\def\RWF{\textrm{\mbox{\tiny{RWF}}}}

\def\º{\textrm{\textordmasculine}}

\begin{document}


\title[LISA parameter estimation of supermassive black holes]{
LISA parameter estimation of supermassive black holes}

\author{Miquel~Trias$^1$ and Alicia~M.~Sintes$^{1, 2}$}
\address{$^1$ Departament de F\'{\i}sica, Universitat de les Illes
Balears, Cra. Valldemossa Km. 7.5, E-07122 Palma de Mallorca, Spain}
\address{$^2$ Max-Planck-Institut f\"ur Gravitationsphysik (Albert-Einstein-Institut),
Am~M\"uhlenberg 1, 14476~Golm, Germany}
\eads{\mailto{miquel.trias@uib.es}, \mailto{sintes@aei.mpg.de}}

\begin{abstract}
We study parameter estimation of supermassive black holes in the range $10^5-10^8\Ms$ by LISA using the inspiral full post-Newtonian gravitational waveforms, and we compare the results with those arising from the commonly used restricted post-Newtonian approximation. The analysis shows that for observations of the last year before merger, the inclusion of the higher harmonics clearly improves the parameter estimation.
We pay special attention to the source location errors and we study the improvement
on the percentage of sources for which we could potentially identify electromagnetic
counterparts. We also show how the additional harmonics can help to mitigate 
the impact of losing laser links during the mission.
\end{abstract}
\date{\today}


\section{Introduction}
\label{sec:intro}
LISA will be an astronomical observatory of unprecedented versatility and range
\cite{Pre-Phase A Report,Danzmann:2003}.
Among the wide range of different LISA sources, the observation of supermassive black holes, in the range $10^5-10^8\Ms$,
merging in galaxies at all distances will address many of LISA's science objectives. 
In particular, they will provide valuable information about the mechanism of their formation \cite{Hughes:2001ya, Menou:2001hb}
and they will also serve as laboratories for fundamental tests of gravitational theory \cite{Dreyer:2003bv,Miller:2004va,
Hughes:2004vw,Berti:2004bd,Arun:2006yw,Arun:2006hn}. 
In addition, 
since many of these supermassive black hole mergers are likely to have electromagnetic counterparts, it will be possible to constrain the values of cosmological parameters by combining the gravitational wave and electromagnetic 
observations \cite{Hughes:2001ya, Schutz:1986gp, Kocsis:2005vv, Kocsis}. 
Moreover, using the distance-redshift relation from many supermassive black holes, LISA will be able to put interesting  constraints on the equation of state of dark energy  \cite{Holz:2005df}.
The real impact of these observations will depend on how accurately the source parameters can be estimated.

Supermassive black hole binaries are long lived sources in the LISA band. The whole coalescence of the compact binary system is usually divided into three phases: the adiabatic inspiral, the merger and the ringdown. Most of the signal-to-noise ratio (SNR) accumulates during the last days of the coalescence, but one critically relies on long integration times to disentangle the source parameters, in particular to resolve the source position in the sky and measure its luminosity distance \cite{Cutler:1997ta}.

In this paper we study parameter estimation of supermassive black hole binary systems in the final stage of inspiral (ignoring the merger phase and the ringdown)
 using the full post-Newtonian gravitational waveforms and we compare the results with those arising from the commonly used restricted post-Newtonian approximation, complementing the results presented in \cite{Arun:2007qv, Arun:2007hu, Trias:2007fp}

The rest of the paper is organized as follows. In section \ref{sec:smbh} we describe the waveform model we  employ and spell out the assumptions on which our analysis is based.  In section \ref{sec:results} we present the results exploring the vast parameter space, paying special attention to the source location errors, and we also compare the results
in the hypothetical case in which LISA could lose two laser links and would operate as a single interferometer. Section \ref{sec:summary} concludes with a summary and pointers to future work.


\section{Observation of supermassive black hole inspiral systems by LISA}
\label{sec:smbh}

The coalescence of binary black holes is commonly divided 
into three successive epochs in the
time domain: inspiral, merger and ringdown. During the inspiral the distance
between the black holes diminishes and the orbital frequency sweeps up.
The waveforms are well modeled using the post-Newtonian
approximation to general relativity (see \cite{Blanchet:2002av} and references therein).
Eventually the post-Newtonian
description of the orbit breaks down, and the black holes cannot be
treated as point particles any more. What is more, it is expected that they
will reach the \emph{innermost stable circular orbit} (ISCO), at which the
gradual inspiral ends and the black holes begin to plunge together to
 form a single black hole.
This is referred as the merger phase. At the end,
 the final black hole will gradually settle down into a Kerr black hole.

The inspiral post-Newtonian (PN) waveforms in the two polarizations 
  $h_+$ and $h_{\times}$,  take the general form
\bea
\label{expansion}
h_{+,\times}&=& {2M\eta\over D_L}(M\omega)^{2/3} \left\{H_{+,\times}^{(0)}
+v^{1/2}H_{+,\times}^{(1/2)}+v H_{+,\times}^{(1)} \right.\nonumber\\
& & \left.
 +v^{3/2}H_{+,\times}^{(3/2)}
+v^2 H_{+,\times}^{(2)}+ v^{5/2}H_{+,\times}^{(5/2)} + \ldots \right\} \ ,
\eea
where we have set  $G=c=1$, 
 $v\equiv (M\omega)^{2/3}$ is the PN expansion factor, $\omega$ is the  orbital frequency,
$D_L$ is the luminosity distance to the source, and $M$ and $\eta$ are the observed total mass
and the symmetric mass ratio respectively.
The explicit expressions for $H_{+,\times}^{(m/2)}$ can be found in
\cite{Blanchet:1996pi,Arun:2004ff} and they include contributions from several harmonics
of the binary's orbital motion.
 Equation~(\ref{expansion}) corresponds to the 
so-called \textit{full} waveform (FWF). 
If one neglects all  amplitude terms
 except the leading Newtonian quadrupole one, but keeping the phase to some specific PN
 order, i.e., keeping only $H_{+,\times}^{(0)}$ and throwing out the rest
  $H_{+,\times}^{(m/2)}$ for  $m>0$, it becomes the \textit{restricted} waveform (RWF).

For the supermassive black hole binary inspirals most of the  SNR accumulates
at frequencies $f<10$~mHz, so it is adequate to use the low-frequency approximation to the LISA response
function derived by Cutler~\cite{Cutler:1997ta}. In this approximation, LISA 
can be regarded as two
independent Michelson interferometers, and 
the strain $h(t)$ produced   by a gravitational wave  signal becomes
\begin{equation}
h^{(i)}(t) = {\sqrt{3} \over 2}\left[F_+^{(i)}(t)h_+(t) + F_\times^{(i)}(t)h_\times(t) \right]
\, ,
\end{equation}
where $F_+^{(i)}$ and $F_\times^{(i)}$ are the time-dependent  antenna 
pattern functions
and
the $i=$I,II labels the two independent Michelson outputs.

The total noise that affects any LISA observation 
 is given by the superposition of instrumental sources, $S_n^\inst(f)$, and astrophysical
 foregrounds of unresolved radiation, $S_n^\conf(f)$.
In this paper we use the same analytical expressions given in~\cite{Barack:2004wc}
\bea
S_n^\inst(f) &= &6.12\times 10^{-51}~f^{-4} + 1.06\times 10^{-40} \nonumber\\
& & +6.12\times 10^{-37}~f^2~ \mathrm{Hz}^{-1} \, .
\eea
For the confusion noise we consider only noise from short-period galactic and extragalactic 
binaries (due to white dwarfs binaries), assuming they are all unresolvable, and we ignore the effects
of captures of compact objects. 

In this paper we employ the Fisher matrix approach to 
study the problem of parameter estimation for supermassive black hole inspirals, paying especial attention to the improvement in errors estimation using the FWF in comparison to the RWF. The analysis performed is based on the following assumptions:
\begin{itemize}
\item We consider the last year of the inspiral phase of the coalescence. We terminate the signal when the binary members are separated by a distance 6$M$ and we also  impose a low-frequency cut-off to the instrument at $5\times 10^{-5}$ Hz.
\item For the waveform model, we restrict ourselves to circular orbits and we take care of spins contributions only in the waveform phase, ignoring spin-induced precession of the orbital plane. Moreover,  we focus our attention to the cases in which spins are negligible.
\item We approximate the waveform at the 2PN order, both in amplitude and phase, considering up to six harmonics in the case of FWF. These waveforms depend on eleven  independent parameters:
\be
{\bm \lambda}=\{ \cos \theta_N, \phi_N ,\cos \theta_L,\phi_L,\ln D_L, 
 t_c, \phi_c, \beta, \sigma, \ln \Mc, \ln \mu \} \, ,
\ee
four angles defining the source position and orientation, the luminosity distance,  $t_c$ and $\phi_c$ are the time and phase at coalescence,  $\beta$ 
and $\sigma$ are the  spin-orbit and spin-spin parameters, and two mass parameters.
\item  We consider sources  at redshift $z=1$ in a flat
Universe described by the cosmological parameters
 $H_0 = 71~\mathrm{km}~\mathrm{s}^{-1}~\mathrm{Mpc}^{-1}$,
 $\Omega_m = 0.27$ and $\Omega_\Lambda = 0.73$.
\end{itemize}

\section{Results }
\label{sec:results}

The results we present here complement and extend those presented in 
\cite{Arun:2007qv, Arun:2007hu, Trias:2007fp}. These papers studied
the effects of adding higher post-Newtonian order corrections to the amplitude, 
analyzing how the SNR and the measurement errors evolve with the total mass
of the system for different values of the mass ratio. Their conclusions were the following: 
\begin{itemize}
\item For the SNR, the contribution of the second harmonic dominates for systems with 
$M < 4\times 10^7 M_{\odot}$, but for higher mass systems the second harmonic is no longer visible and then  the contribution of the higher harmonics becomes relevant, increasing LISA's mass reach for supermassive black holes. 
\item For the measurement errors, the use of higher  harmonic terms
improves the parameter estimation, not only for the most massive systems, but for any binary with total mass higher than
$10^5 M_{\odot}$. The angular resolution has gain factors from $2-3$ up
to one order of magnitude when they go to higher masses, and the  estimation of masses has even better
improvements, with gain factors of at least one order of magnitude.
\end{itemize}

\begin{table*}
\caption{$10 \%$, $50 \%$ and $90 \%$ levels of the cumulative probability distributions of SNR and
measurement errors for different  pair of masses using FWF and RWF, together with the
gain factors computed as 
$\langle SNR_{50 \%}\rangle_{\FWF}/ \langle SNR_{50 \%}\rangle_{\RWF}$
and  $ 10^{(\langle x_{50 \%}\rangle_{\RWF}-\langle x_{50 \%}\rangle_{\FWF})}$, for the SNR and measurement errors,
respectively.
The probability distributions are obtained by considering LISA as a combination of two independent Michelson interferometers.
}
\begin{center}
\begin{tabular}{lccccccc}
\hline 
& \multicolumn{3}{c}{RWF} & \multicolumn{3}{c}{FWF}  & Gain \\
$x$  &  $x_{10 \%}$ & $x_{50 \%}$ & $x_{90 \%} \quad$ &  $\quad x_{10 \%}$ & $x_{50 \%}$ & $x_{90 \%}$  & $~$ factor $~$\\

\hline
\\
& \multicolumn{7}{c}{\begin{normalsize} (a) $m_1 = 10^7 M_{\odot}$~;~$m_2 = 10^7 M_{\odot}$\end{normalsize}} \\

 SNR   & $174$ &  $333$ & $636$   &   $152$ &  $287$   &  $558$   &   $0.86$ \\
 $\log_{10} \Delta\Omega_N /$srad   & $-1.48$ &  $-0.20$  &  $0.55$    &   $-4.16$ & $-1.78$  & $0.48$   &   $38$   \\
 $\log_{10} \Delta\Omega_L / $srad   & $-0.70$ &  $0.17$  &  $1.65$    &   $-2.14$ & $-1.44$  & $-0.07$   &   $40$   \\
 $\log_{10} \Delta D_L / D_L$   & $0.66$  & $0.95$  &  $1.22$     &   $-1.29$ & $-0.93$ & $-0.27$   &   $76$   \\
 $\log_{10} \Delta t_c / $s   & $4.98$ & $5.27$  &  $5.54$    &    $3.18$  & $3.42$ & $3.65$   &   $71$  \\
 $\log_{10} \Delta \cm / \cm$   & $0.72$ & $1.01$  & $1.28$    &    $-1.67$ & $-1.47$  & $-1.25$   &   $300$   \\
 $\log_{10} \Delta \mu / \mu$   & $2.55$ &  $2.84$ & $3.11$    &    $-1.67$ & $-1.47$  &  $-1.25$   &   $20000$  \\

\\
& \multicolumn{7}{c}{\begin{normalsize} (b) $m_1 = 10^7 M_{\odot}$~;~$m_2 = 10^6 M_{\odot}$\end{normalsize}} \\

 SNR   & $121$ &  $231$ & $445$   &   $116$ &  $212$   &  $403$   &   $0.92$  \\
 $\log_{10} \Delta\Omega_N /$srad   & $-2.74$ &  $-1.47$  &  $-0.73$    &   $-4.03$ & $-2.46$  & $-1.14$   &   $9.7$  \\
 $\log_{10} \Delta\Omega_L / $srad   & $-1.93$ &  $-1.14$   &  $0.43$   &   $-2.88$ & $-2.39$  & $-1.55$   &   $18$  \\
 $\log_{10} \Delta D_L / D_L$   & $-0.88$  & $-0.60$  &  $-0.27$     &   $-1.74$ & $-1.41$ & $-1.02$   &   $6.4$  \\
 $\log_{10} \Delta t_c / $s   & $3.81$ &  $4.10$  &  $4.37$    &    $2.66$  & $2.84$ & $2.98$   &   $18$  \\
 $\log_{10} \Delta \cm / \cm$   & $-1.00$ &  $-0.71$ & $-0.44$    &    $-2.79$ & $-2.66$  & $-2.50$   &   $88$  \\
 $\log_{10} \Delta \mu / \mu$   & $0.71$ &  $1.00$ & $1.28$    &    $-2.17$ & $-1.79$  &  $-1.40$   &   $610$  \\

\\
& \multicolumn{7}{c}{\begin{normalsize}(c) $m_1 = 10^6 M_{\odot}$~;~$m_2 = 10^6 M_{\odot}$\end{normalsize}} \\

 SNR   & $194$ &  $362$ & $698$   &   $171$ &  $328$   &  $632$   &   $0.90$  \\
 $\log_{10} \Delta\Omega_N /$srad   & $-3.57$ &  $-2.42$  &  $-1.84$    &   $-4.35$ & $-2.72$  & $-1.85$   &   $2.0$  \\
 $\log_{10} \Delta\Omega_L / $srad   & $-2.86$ &  $-2.15$   &  $-0.54$   &   $-2.90$ & $-2.38$  & $-2.05$   &   $1.7$  \\
 $\log_{10} \Delta D_L / D_L$   & $-1.78$  & $-1.37$  &  $-0.89$     &   $-1.93$ & $-1.58$ & $-1.30$   &   $1.6$  \\
 $\log_{10} \Delta t_c / $s   & $1.59$ &  $1.82$  &  $2.07$    &    $1.13$  & $1.38$ & $1.62$   &   $2.8$  \\
 $\log_{10} \Delta \cm / \cm$   & $-2.92$ &  $-2.66$ & $-2.42$    &    $-3.85$ & $-3.63$  & $-3.39$   &   $9.3$  \\
 $\log_{10} \Delta \mu / \mu$   & $-0.73$ &  $-0.48$ & $-0.23$    &    $-3.85$ & $-3.63$  &  $-3.39$   &   $1400$  \\

\\
& \multicolumn{7}{c}{\begin{normalsize}(d) $m_1 = 10^6 M_{\odot}$~;~$m_2 = 10^5 M_{\odot}$\end{normalsize}} \\

 SNR   & $163$ &  $311$ & $597$   &   $187$ &  $322$   &  $578$   &   $1.0$   \\
 $\log_{10} \Delta\Omega_N /$srad   & $-3.57$ &  $-2.51$  &  $-1.99$    &   $-4.40$ & $-3.00$  & $-2.02$   &   $3.0$  \\
 $\log_{10} \Delta\Omega_L / $srad   & $-2.91$ &  $-2.23$  &  $-0.42$    &   $-3.52$ & $-2.99$  & $-2.45$   &   $5.9$  \\
 $\log_{10} \Delta D_L / D_L$   & $-1.81$  & $-1.41$  &  $-0.87$     &   $-2.13$ & $-1.79$ & $-1.38$   &   $2.4$ \\
 $\log_{10} \Delta t_c / $s   & $1.32$ & $1.56$  &  $1.80$    &    $0.90$  & $1.10$ & $1.32$   &   $2.9$  \\
 $\log_{10} \Delta \cm / \cm$   & $-3.49$ & $-3.27$  & $-3.05$    &    $-4.26$ & $-4.05$  & $-3.86$   &   $6.0$  \\
 $\log_{10} \Delta \mu / \mu$   & $-1.46$ &  $-1.24$ & $-1.01$    &    $-2.51$ & $-2.25$  &  $-1.91$   &   $10$  \\

\hline
\end{tabular}
\label{Tab.MCs}
\end{center}
\end{table*}

In this paper we analyze in more detail the effects of the higher harmonics corrections
in the distributions of the parameter estimation errors, interpreting the results in a complete new way that provides more useful information about the impact of using the higher harmonics in modeling supermassive black holes. In particular we pay special attention to the source location errors and we study the improvement on the percentage of sources for which we could potentially identify electromagnetic counterparts. 
We also study how the additional harmonics in the FWF can help to mitigate the impact of losing laser links during the mission, therefore extending the previous results.

For this analysis, we perform extensive  Monte-Carlo simulations
in order to extract general conclusions of which is the real impact of using these 
higher harmonics corrections, since the results can vary significantly from source to source.
As we did in a previous paper \cite{Trias:2007fp},
 for different pair of masses, we consider an ensemble of $1000$ fiducial sources all at 
redshift $z=1$ (which, according to our Universe model, sets the luminosity distance to
$D_L = 6.64~\Gpc$), with zero spins $\beta = \sigma = 0$, and we select randomly
the four geometrical angles ($\theta_N$, $\phi_N$, $\theta_L$ and $\phi_L$) from an uniform
distribution in $\cos \theta_N$, $\phi_N$, $\cos \theta_L$ and $\phi_L$.  

The probability distributions of SNR and measurements errors for observations of the final year of 
supermassive black hole binaries, by considering LISA as a two independent Michelson interferometers, can be found in figure 11 of \cite{Trias:2007fp}. 
Since LISA's measurement errors span several orders of magnitude 
with just changing the sky location and orientation of the source
(see e.g. figures 7--10 of \cite{Trias:2007fp}),
it is very important to 
characterize  these distributions properly. For this reason,  we provide
 in table \ref{Tab.MCs} the $10~\%$, $50~\%$ and $90~\%$ levels of cumulative probability distributions, for the SNR and the measurement errors,  for four particular pair of masses
using the RWF and the FWF, together with the gain factors.
The results presented in table \ref{Tab.MCs} correspond again to the most interesting case in 
which LISA has all its six laser links working.
Notice that for all the parameters, the median values in the case of the 
FWF distributions show an improvement with respect to the 
RWF ones, which for some of them is more than one order of magnitude.
The difference between the  $10~\%$ and $90~\%$ levels of cumulative probability distributions
provide the error intervals for $80~\%$ of the source population.

From table \ref{Tab.MCs}, the reader can see the remarkable improvements for the 
$10^7 M_{\odot}-10^7 M_{\odot}$, and $10^7 M_{\odot}- 10^6 M_{\odot}$ cases in 
angular resolution and  distance measurement:
the angular resolution improves the median value by factors of
$38$ and $9.7$, respectively; and  the luminosity distance by $76$ and $6.4$,
respectively. One should also notice that, in these two cases, 
 those parameters were poorly determined 
using only the RWF. For the other sets of masses the averaged improvement in
angular resolution and luminosity distance is more moderate, 
between $2-3$ for $\Delta\Omega_N$,
and $1.6-2.4$ for $\Delta D_L/D_L$.
In all cases the masses are determined much more accurate, even by several 
orders of magnitude in the case of $\mu$, using the FWF. For the equal mass cases,
 the errors in $\Mc$ and $\mu$ are of
 the same order using the FWF.

\begin{table*}
\caption{Percentage of sources from Monte Carlo simulations of table \ref{Tab.MCs} with a measurement error
less than the given values.}
\begin{center}
\begin{tabular}{lcccccccc}
\hline 
\multicolumn{1}{r}{$m_1 , m_2 (M_{\odot})$:} & \multicolumn{2}{c}{$10^7 - 10^7$} & \multicolumn{2}{c}{$10^7 - 10^6$}
& \multicolumn{2}{c}{$10^6 - 10^6$}  & \multicolumn{2}{c}{$10^6 - 10^5$} \\
&   \textsc{rwf} & \textsc{fwf}  $~$ &   \textsc{rwf} & \textsc{fwf} $~$  &   \textsc{rwf} & \textsc{fwf} $~$  &   \textsc{rwf} & \textsc{fwf}   \\
\hline
$\Delta\Omega_N  < (2.5\º \times 2.5\º)$    &   $2.4$ & $34.2~$   &   $10$ & $43~$   &
                                                                   $35$ & $50~$   &   $35$ & $57$   \\
$\Delta\Omega_N  < (1\º \times 1\º)$    &   $0.8$ & $19.3~$   &   $3.1$ & $24.1~$   &  
                                                              $11$ & $28~$   &   $10$ & $35$  \\
$\Delta\Omega_N  < (0.5\º \times 0.5\º)$    &   $0.4$ & $10.7~$   &   $1.0$ & $7.4~$   &
                                                                   $3.9$ & $14.0~$   &   $3.5$ & $18.5$  \\
$\Delta D_L / D_L  < 10 \%$    &   $0$ & $42~$   &   $2$ & $92~$   & 
                                                  $86$ & $100~$   &   $85$ & $100$   \\
$\Delta \cm / \cm  < 10 \%$    &   $0$ & $99~$   &   $10$ & $100~$   & 
                                                $100$ & $100~$   &   $100$ & $100$   \\
$\Delta \mu / \mu  < 10 \%$    &   $0$ & $99~$   &   $0$ & $99~$   &
                                                 $0$ & $100~$   &  $91$ & $100$   \\
\hline
\end{tabular}
\label{Tab.MCs2}
\end{center}
\end{table*}

Another interesting way of looking at these probability distributions, is computing the percentage
of fiducial sources that would have an error less than a certain number, and compare how
 these populations change when we use one or other waveform model. In table \ref{Tab.MCs2}
we summarize these quantities for some of the most interesting parameters.
In our analysis we pay special attention to the improvement of LISA angular resolution 
due to the inclusion of the higher harmonics in the waveform. One of LISA's objectives 
is to detect the signal during the 
inspiral phase alone and estimate its sky location accurately enough in order to issue warnings to the astronomical 
community of possible simultaneous observation of merger (and ringdown) with X-ray and optical observatories. The aim would be
to identify the electromagnetic counterpart (host galaxy  or galaxy cluster) and be able to disentangle the redshift parameter, since
exiting science can be done by studying the luminosity distance redshift relation 
\cite{Schutz:1986gp, Kocsis:2005vv, Kocsis, Holz:2005df, Arun:2007hu}.

Taking into account real field of view values of some of the 
existing or planned wide field electromagnetic instruments (see table 1 of \cite{arxiv:0712.1144}),
we have studied the number of sources that could be observed by LISA with an angular
resolution better than $(2.5\º \times 2.5\º)$, $(1\º \times 1\º)$  and $(0.5\º \times 0.5\º)$ corresponding 
to $1.9\times 10^{-3}$, $3.0\times 10^{-4}$  and $7.6\times 10^{-5}$ srad, respectively. 
 For the mass cases considered in table \ref{Tab.MCs2}, we have that using the FWF at least $20~\%$ of the binary
systems, located at $z=1$ observed by LISA during the last year of coalescence, will have an
angular resolution better than ($1\º \times 1\º$). This represents $15~\%$ more of sources 
than if we were using the RWF. Moreover, 
 except for the very high mass systems $M>10^7~M_{\odot}$, we would have at least $14~\%$ of sources resolved with
an angular resolution better than $0.5\º \times 0.5\º$
if we worked with FWF, while these numbers drop to less than $4~\%$ for the RWF case. Therefore, we can conclude that the 
inclusion of the  higher harmonics in the waveform is crucial in terms of being able or not to make
observations of their electromagnetic counterpart.
Concerning the other parameters in table \ref{Tab.MCs2}, luminosity distance and masses,
the errors are less than $10~\%$ in almost all the cases using the FWF, while 
this is not the case with the RWF model. Notice that results depend a lot on the masses of the binary.

\begin{figure}
\includegraphics[width=\textwidth]{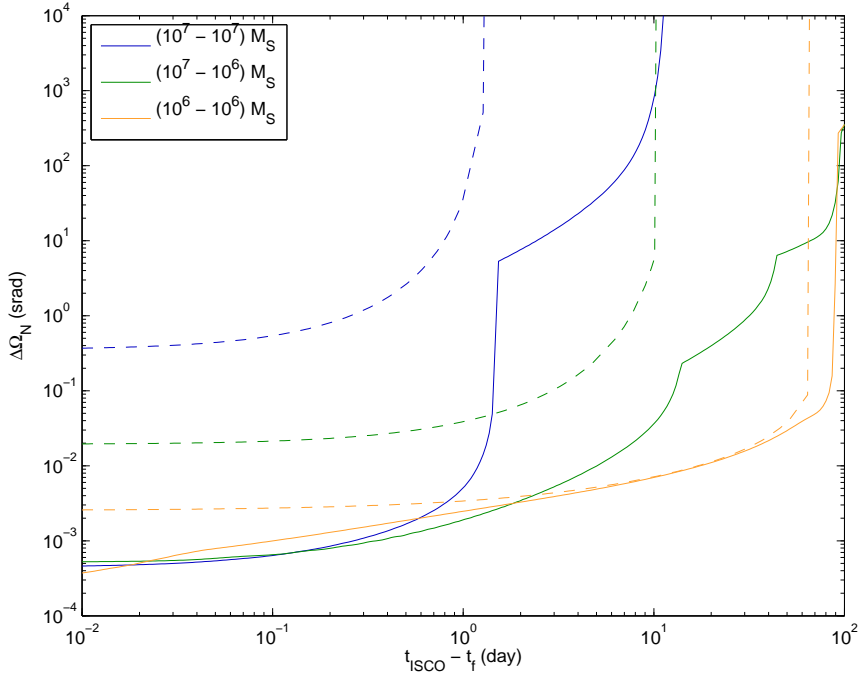} 
\caption{Time evolution of LISA's angular resolution when we observe different supermassive
black hole binary systems from one year before the merger up to $t_f$. Lookback time is represented
the axis of abscissas, so time evolves from right to left. Fiducial sources are located at redshift $z=1$, with a
common orientation defined by $\cos \theta_L = 0.2$ and $\phi_L = 3$ and a sky location of
 $\cos \theta_N = -0.383$ ; $\phi_N = 2.82$.
Solid lines correspond to FWF, and the dashed ones to RWF.}
\label{Fig.Sirens}
\end{figure}

In order to issue warnings and search for electromagnetic counterparts, we need LISA to have enough
angular resolution some time before the  merger occurs.  For this reason,
we are not only interested in LISA's angular resolution when
we observe a source during the last year before merger, but also how the angular resolution evolves when
we stop our observation some time before, and measure the importance of the FWF versus RWF.
%
In figure \ref{Fig.Sirens} we study how the angular resolution is accumulated as a function of the
lookback time, $t_{ISCO} - t_f$,  for different sources located at 
$\cos \theta_N = -0.383$, $\phi_N = 2.82$. 
Different sky  locations have also been analyzed, see e.g. figure 14 in \cite{Trias:2007fp} for 
$\cos \theta_N = -0.6$, $\phi_N = 1$. In this latter figure,
we have that an angular resolution twice the final one is reached, respectively, $20$ days, $20$ hours and 
$8$ hours before merger, for binary systems of masses  $(10^6 - 10^6)~M_{\odot}$, $(10^7 - 10^6)~M_{\odot}$ and 
$(10^7 - 10^7)~M_{\odot}$.  For the sky location considered here
(figure \ref{Fig.Sirens}),  LISA's
angular resolution decreases more during the last days (hours) before merger, but we still are 
reaching almost the best angular resolution $12$ hours before merger.
The jumps in the evolution of LISA's angular resolution correspond 
to those times in which a new harmonic enters into the LISA band and it is related to the 
the low-frequency cut-off we have assumed for the instrument at  $5\times 10^{-5}$ Hz.
For equal masses, only the even multipoles of the orbital frequency contribute to the FWF, while for the unequal masses 
there is contribution from all the harmonics.
For example, in the $10^7 M_{\odot}-10^7 M_{\odot}$ case the contribution 
of the 4th harmonic becomes relevant
around 10 days before coalescence (independently of the sky location considered)
 while the 2nd harmonic rapidly increases the angular resolution 2 days before
coalescence. For the unequal mass case $10^7 M_{\odot}- 10^6 M_{\odot}$ we clearly see the
contributions of the 2nd, 3rd and 4th harmonics.

%
%
%
\begin{table*}
\caption{Comparison between median values of the probability distributions when we see
LISA as a single Michelson interferometer or as a combination of two independent ones. Gain
factors are computed in the same way as in table~\ref{Tab.MCs} and by \emph{loss}
we mean the factor that the errors get multiplied by (and the SNR divided by) when we lose
one of the LISA's arms.}
\begin{center}
\begin{tabular}{lcccccccc}
\hline 
& \multicolumn{3}{c}{Detector $I$} & \multicolumn{3}{c}{Detector $I+II$}  & \multicolumn{2}{c}{Loss} \\
& \multicolumn{2}{c}{$x_{50 \%}$} &    & \multicolumn{2}{c}{$\quad x_{50 \%}$} &   &  &  \\
$x$   & \textsc{rwf} & \textsc{fwf} & Gain $\quad$ & \textsc{rwf} & \textsc{fwf} & Gain $\quad$  &  $\quad$ \textsc{rwf} & \textsc{fwf}  \\

\hline
\\
& \multicolumn{7}{c}{\begin{normalsize} (a) $m_1 = 10^7 M_{\odot}$~;~$m_2 = 10^6 M_{\odot}$\end{normalsize}} \\

 SNR   & $161$ &  $152$ & $0.94$   &   $231$ &  $212$   &  $0.92$   &   $1.4$  &  $1.4$  \\
 $\log_{10} \Delta\Omega_N /$srad   & $3.40$ &  $-0.69$  &  $12000$    &   $-1.47$ & $-2.46$  & $9.7$   &   $74000$  &  $59$  \\
 $\log_{10} \Delta\Omega_L / $srad   & $3.95$ &  $-0.89$   &  $69000$   &   $-1.14$ & $-2.39$  & $18$   &   $120000$  &  $32$  \\
 $\log_{10} \Delta D_L / D_L$   & $1.84$  & $-0.69$  &  $330$     &   $-0.60$ & $-1.41$ & $6.4$   &   $276$  &  $5.2$  \\
 $\log_{10} \Delta t_c / $s   & $4.68$ &  $3.02$  &  $46$    &    $4.10$  & $2.84$ & $18$   &   $3.9$  &  $1.5$  \\
 $\log_{10} \Delta \cm / \cm$   & $0.05$ &  $-2.46$ & $320$    &    $-0.71$ & $-2.66$  & $88$   &   $5.8$  &  $1.6$  \\
 $\log_{10} \Delta \mu / \mu$   & $1.70$ &  $-1.52$ & $1700$    &    $1.00$ & $-1.79$  &  $610$   &   $5.0$  &  $1.8$  \\

\\
& \multicolumn{7}{c}{\begin{normalsize}(b) $m_1 = 10^6 M_{\odot}$~;~$m_2 = 10^6 M_{\odot}$\end{normalsize}} \\

 SNR   & $257$ &  $229$ & $0.89$   &   $362$ &  $328$   &  $0.90$   &   $1.4$  &  $1.4$  \\
 $\log_{10} \Delta\Omega_N /$srad   & $-0.96$ &  $-1.56$  &  $4.0$    &   $-2.42$ & $-2.72$  & $2.0$   &   $29$  &  $14$  \\
 $\log_{10} \Delta\Omega_L / $srad   & $-0.19$ &  $-1.47$   &  $19$   &   $-2.15$ & $-2.38$  & $1.7$   &   $92$  &  $8.0$  \\
 $\log_{10} \Delta D_L / D_L$   & $-0.39$  & $-1.05$  &  $4.6$     &   $-1.37$ & $-1.58$ & $1.6$   &   $9.6$  &  $3.4$  \\
 $\log_{10} \Delta t_c / $s   & $2.16$ &  $1.70$  &  $2.9$    &    $1.82$  & $1.38$ & $2.8$   &   $2.2$  &  $2.1$  \\
 $\log_{10} \Delta \cm / \cm$   & $-2.30$ &  $-3.36$ & $12$    &    $-2.66$ & $-3.63$  & $9.3$   &   $2.3$  &  $1.9$  \\
 $\log_{10} \Delta \mu / \mu$   & $-0.14$ &  $-3.36$ & $1700$    &    $-0.48$ & $-3.63$  &  $1400$   &   $2.2$  &  $1.9$  \\

\\
& \multicolumn{7}{c}{\begin{normalsize}(c) $m_1 = 10^6 M_{\odot}$~;~$m_2 = 10^5 M_{\odot}$\end{normalsize}} \\

 SNR   & $219$ &  $235$ & $1.1$   &   $311$ &  $322$   &  $1.0$   &   $1.4$  &  $1.4$   \\
 $\log_{10} \Delta\Omega_N /$srad   & $-1.75$ &  $-1.97$  &  $1.7$    &   $-2.51$ & $-3.00$  & $3.0$   &   $5.8$  &  $11$  \\
 $\log_{10} \Delta\Omega_L / $srad   & $-0.96$ &  $-2.08$  &  $13$    &   $-2.23$ & $-2.99$  & $5.9$   &   $18$  &  $8.2$  \\
 $\log_{10} \Delta D_L / D_L$   & $-0.79$  & $-1.36$  &  $3.7$     &   $-1.41$ & $-1.79$ & $2.4$   &   $4.2$  &  $2.7$ \\
 $\log_{10} \Delta t_c / $s   & $1.76$ & $1.38$  &  $2.4$    &    $1.56$  & $1.10$ & $2.9$   &   $1.6$  &  $1.9$  \\
 $\log_{10} \Delta \cm / \cm$   & $-3.06$ & $-3.86$  & $6.3$    &    $-3.27$ & $-4.05$  & $6.0$   &   $1.6$  &  $1.6$  \\
 $\log_{10} \Delta \mu / \mu$   & $-1.03$ &  $-2.02$ & $10$    &    $-1.24$ & $-2.25$  &  $10$   &   $1.6$  &  $1.7$  \\

\hline
\end{tabular}
\label{Tab.TwoDets}
\end{center}
\end{table*}

Up to this point, we have considered LISA as two independent Michelson interferometers, using the 
 long wavelength approximation, and we have provided the measurement errors for this two-detector case. But
it  is also interesting to study which is the impact on parameter estimation for a single interferometer.
 This is relevant in the hypothetical  case that LISA would lose one of its arm links, in which case
 it would be interesting to know how LISA's parameter estimation gets
reduced and whether the effect is the same when we use the RWF model or the FWF one.

In table \ref{Tab.TwoDets}, we give the $50~\%$ levels of the  cumulative probability distributions working with
both RWF and FWF, for the single detector  and the two-detector case.
 As one would expect, each of the independent interferometers measures,  in average, similar values of SNR, and since 
$\rho^{tot} = \sqrt{(\rho^I)^2 + (\rho^{II})^2}$,  losing one of LISA's link means
that the SNR gets reduced by a factor $\sqrt{2}$ in all cases. For the measurement errors, we 
see that in most of the cases (especially for high mass systems) the loss 
factor (\ie the factor that errors have to be multiplied by when LISA loses one of its links) is higher for the 
RWF case, which means that working with the FWF not only improves LISA's parameter estimation as 
we have seen before, but also reduces the impact of losing one arm link.  
In any case, even working with the FWF model, we see that losing one arm, means reducing the
angular resolution by more than one order of magnitude, thus we will lose any possibility of observing
the potential electromagnetic counterpart of the merger. Losses in the other parameters are not so serious:
errors in the distance get multiplied by a factor $2-5$ and in masses by less than $2$.


\section{Summary and outlook}
\label{sec:summary}

This work clearly shows that modeling the inspiral with the full post-Newtonian waveform, as compared to the restricted-PN one, improves in general the parameter estimation of supermassive black hole systems, 
due to the much greater richness of the waveform.
One should notice that the results presented here are influenced by a number of assumptions associated to these observations, in particular, the instrumental and confusion noise models we have adopted, the post-Newtonian waveform model and how this is terminated, and the lower-frequency cut-off we have imposed. The parameter estimation code we use has been validated by the 
\textit{LISA science performance evaluation taskforce} \cite{lisape}. This new taskforce, created at the September'07 LISA International Science Team meeting, is currently extending these results to include spin-induced precession into the full waveform and compare LISA's science reach for different mission configurations, based also on  astrophysically motivated source distributions \cite{Berti:2008af}.
It will be very interesting to extend these analyses using different post-Newtonian models and also to study parameter estimation
for all three stages of the signal, including the merger and the ringdown, and use currently available phenomenological and numerical waveforms for that.

\section*{Acknowledgments}

We would like to thank K.~G.~Arun, S.~Babak, C.~V.~D.~Broeck, N.~Cornish, C.~Cutler, D.~Holz, S.~A.~Hughes, S.~Husa,  B.~R.~Iyer,  
B.~Kocsis, E.~Porter,  B.~S.~Sathyaprakash,  B.~F.~Schutz,  A.~Vecchio and the LISA science performance evaluation
taskforce for many helpful discussions.
  We also acknowledge the support of the Max-Planck
Society, the Spanish  Ministerio de Educaci\'on y Ciencia
Research Projects FPA-2007-60220, HA2007-0042, CSD207-00042 and the Govern de les Illes
Balears, Conselleria d'Economia, Hisenda i Innovaci\'o.  
We are also grateful to the Albert
Einstein Institute for hospitality where this work was initiated.
 

\end{document}